\documentclass{article}
\usepackage{ijcai18}

\usepackage{times}
\usepackage{xcolor}
\usepackage{soul}
\usepackage[utf8]{inputenc}
\usepackage[small]{caption}
\usepackage{url}  %Required
\usepackage{graphicx}  %Required
\usepackage{subfig}
\usepackage{bm}
\usepackage{tabularx}                             % more flexible table environment
\usepackage{algorithm}
\usepackage{algorithmicx}
\usepackage{amssymb}
\usepackage[noend]{algpseudocode}
\usepackage{multicol}
\usepackage{amsmath}
\newcolumntype{P}[1]{>{\centering\arraybackslash}p{#1}}

\title{Leveraging Statistical Multi-Agent Online Planning with Emergent Value Function Approximation}
\author{Thomy Phan, Lenz Belzner, Thomas Gabor and Kyrill Schmid\\
Institute of Informatics\\
LMU Munich\\
\{thomy.phan, belzner, thomas.gabor, kyrill.schmid\}@ifi.lmu.de}

\begin{document}

\maketitle

\begin{abstract}
Making decisions is a great challenge in distributed autonomous environments due to enormous state spaces and uncertainty. Many online planning algorithms rely on statistical sampling to avoid searching the whole state space, while still being able to make acceptable decisions. However, planning often has to be performed under strict computational constraints making online planning in multi-agent systems highly limited, which could lead to poor system performance, especially in stochastic domains.
In this paper, we propose \emph{Emergent Value function Approximation for Distributed Environments (EVADE)}, an approach to integrate global experience into multi-agent online planning in stochastic domains to consider global effects during local planning. For this purpose, a value function is approximated online based on the emergent system behaviour by using methods of reinforcement learning.
We empirically evaluated EVADE with two statistical multi-agent online planning algorithms in a highly complex and stochastic smart factory environment, where multiple agents need to process various items at a shared set of machines. Our experiments show that EVADE can effectively improve the performance of multi-agent online planning while offering efficiency w.r.t. the breadth and depth of the planning process.
\end{abstract}

\section{Introduction}\label{sec:introduction}

Decision making in complex and stochastic domains has been a major challenge in artificial intelligence for many decades due to intractable state spaces and uncertainty. Statistical approaches based on Monte-Carlo methods have become popular for planning under uncertainty by guiding the search for policies to more promising regions in the search space \cite{kocsis2006bandit,silver2010monte,weinstein2013open,amato2015scalable,belzner2015onplan,silver2016mastering,claes2017decentralised}. These methods can be combined with online planning to adapt to unexpected changes in the environment by interleaving planning and execution of actions \cite{silver2010monte,amato2015scalable,belzner2015onplan,silver2016mastering,claes2017decentralised}.

However, online planning often has to meet strict real-time constraints limiting the planning process to local search. This makes the consideration of possible global effects difficult, which could lead to suboptimal policies, especially in stochastic domains. The problem is further intensified in multi-agent systems (MAS), where the search space grows exponentially w.r.t. the dimension and the number of agents, which is known as the \emph{curse of dimensionality} \cite{boutilier1996planning,amato2015scalable,oliehoek2016concise}. Furthermore, one has to cope with the coordination of individual actions of all agents to avoid potential conflicts or suboptimal behaviour \cite{boutilier1996planning,bucsoniu2010multi}.

Many multi-agent planning approaches assume the availability of a pre-computed value function of a more simplified model of the actual environment to consider possible global effects in the local planning process, which can be exploited to prune the search space or to further refine the policy \cite{emery2004approximate,szer2005maa,oliehoek2008optimal,spaan2011scaling}. This might be insufficient for highly complex and uncertain domains, where the dynamics cannot be sufficiently specified beforehand \cite{belzner2015onplan}. Depending on the domain complexity, pre-computing such a value function might be even computationally infeasible \cite{boutilier1996planning,silver2010monte}. Thus, an adaptive and model-free approach is desirable for learning a value function at system runtime in MAS.

Recently, approaches to combine online planning and reinforcement learning (RL) have become popular to play games with high complexity like Go and Hex \cite{silver2016mastering,silver2017mastering,thinkingfastandslow}. A tree search algorithm is used for planning, which is guided by a value function approximated with RL. These approaches were shown to outperform plain planning and RL, even achieving super-human level performance in Go without any prior knowledge about the game beyond its rules \cite{silver2017mastering}. So far, these approaches have only been applied to deterministic domains with only one agent.

In this paper, we propose \emph{Emergent Value function Approximation for Distributed Environments (EVADE)}, an approach to integrate global experience into multi-agent online planning in stochastic domains. For this purpose, a value function is approximated online based on the emergent system behaviour by using methods of RL. With that value function, global effects can be considered during local planning to improve the performance and efficiency of existing multi-agent online planning algorithms.
We also introduce a smart factory environment, where multiple agents need to process various items with different tasks at a shared set of machines in an automated and self-organizing way. Given a sufficient number of agents and stochasticity w.r.t. the outcome of actions and the behaviour of agents, we show that our environment has a significantly higher branching factor than the game of Go \cite{silver2016mastering}.
We empirically evaluate the effectiveness of EVADE in this stochastic and complex domain based on two existing multi-agent planning algorithms \cite{oliehoek2008cross,belzner2017scalable}.

The rest of the paper is organized as follows. Section \ref{sec:background} provides some background about decision making in general. Section \ref{sec:related_work} discusses related work. Section \ref{sec:evade} describes EVADE for enhancing multi-agent planning algorithms. Section \ref{sec:experiments} presents and discusses experimental results achieved by two statistical multi-agent planning algorithms enhanced with EVADE in our smart factory environment. Finally, section \ref{sec:conlusion} concludes and outlines a possible direction for future work.

\section{Background}\label{sec:background}
\subsection{Markov Decision Processes}
We formulate our problem as \emph{multi-agent Markov Decision Process (MMDP)} assuming a fully cooperative setting, where all agents share the same common goal \cite{boutilier1996planning,oliehoek2016concise}. For simplicity, this work only focuses on fully observable problems as modeled in \cite{boutilier1996planning,tampuu2017multiagent,claes2017decentralised}.

Although more realistic models exist for describing large-scale MAS like Dec-MDPs or Dec-POMDPs \cite{oliehoek2016concise}, the focus of this work is just to evaluate the possible performance and efficiency gain based on integrating global experience into the multi-agent online planning process. An extension of our approach to partially observable models is left for future work.

\subsubsection{MDP}
Decision-making problems with discrete time steps and a single agent can be formulated as \emph{Markov Decision Process (MDP)} \cite{beranek1961ronald,boutilier1996planning,puterman2014markov}. An MDP is defined by a tuple $M = \langle\mathcal{S},\mathcal{A},\mathcal{P},\mathcal{R}\rangle$, where $\mathcal{S}$ is a (finite) set of states, $\mathcal{A}$ is the (finite) set of actions, $\mathcal{P}(s_{t+1}|s_{t}, a_{t})$ is the transition probability function and $\mathcal{R}(s_{t}, a_{t})$ is the scalar reward function. In this work, it is always assumed that $s_{t}, s_{t+1} \in \mathcal{S}$, $a_{t} \in \mathcal{A}$, $r_{t} = \mathcal{R}(s_{t}, a_{t})$, where $s_{t+1}$ is reached after executing $a_{t}$ in $s_{t}$ at time step $t$. $\Pi$ is the policy space and $|\Pi|$ is the number of all possible policies.

The goal is to find a \emph{policy} $\pi : \mathcal{S} \rightarrow \mathcal{A}$ with $\pi \in \Pi$, which maximizes the (discounted) return $G_{t}$ at state $s_{t}$ for a horizon $h$:
\begin{equation}\label{eq:return}
G_{t} = \sum_{k=0}^{h-1} \gamma^{k} \cdot \mathcal{R}(s_{t+k}, a_{t+k})
\end{equation}
where $\gamma \in [0,1]$ is the discount factor. If $\gamma < 1$, then present rewards are weighted more than future rewards.

A policy $\pi$ can be evaluated with a \emph{state value function} $V^{\pi} = \mathbb{E}_{\pi}[G_{t}|s_{t}]$, which is defined by the expected return at state $s_{t}$ \cite{bellman1957,beranek1961ronald,boutilier1996planning}. $\pi$ is optimal if $V^{\pi}(s_{t}) \geq V^{\pi'}(s_{t})$ for all $s_{t} \in S$ and all policies $\pi' \in \Pi$. The optimal value function, which is the value function for any optimal policy $\pi^{*}$, is denoted as $V^{*}$ and defined by \cite{bellman1957,boutilier1996planning}:
\begin{equation}\label{eq:policy_value_function}
V^{*}(s_{t}) = max_{a_{t} \in \mathcal{A}}\big\{r_{t} + \gamma \sum_{s' \in \mathcal{S}}^{}P(s'|s_{t},a_{t}) \cdot V^{*}(s')\big\}
\end{equation}

\subsubsection{Multi-Agent MDP}\label{subsubsec:MMDP}
An MMDP is defined by a tuple $M = \langle\mathcal{D},\mathcal{S},\mathcal{A},\mathcal{P},\mathcal{R}\rangle$, where $\mathcal{D} = \{1,...,n\}$ is a (finite) set of agents and $\mathcal{A} = \mathcal{A}_{1} \times ... \times \mathcal{A}_{n}$ is the (finite) set of joint actions. $\mathcal{S}$, $\mathcal{P}$ and $\mathcal{R}$ are defined analogously to an MDP, given joint actions instead of atomic actions \cite{boutilier1996planning}.

The goal is to find a \emph{joint policy} $\pi = \langle\pi_{1},...,\pi_{n}\rangle$, which maximizes the return $G_{t}$ of eq. \ref{eq:return}. $\pi_{i}$ is the individual policy of agent $i \in \mathcal{D}$. Given $n$ agents in the MMDP, the number of possible joint policies is defined by $|\Pi| = \prod_{i=1}^{n}|\Pi_{i}|$. If all agents share the same individual policy space $\Pi_{i}$, then $|\Pi| = |\Pi_{i}|^{n}$.

Similarly to MDPs, a value function $V^{\pi}$ can be used to evaluate the joint policy $\pi$.

\subsection{Planning}
\begin{figure}[!ht]
     \subfloat[local planning\label{fig:open_loop_planning}]{%
       \includegraphics[width=0.23\textwidth]{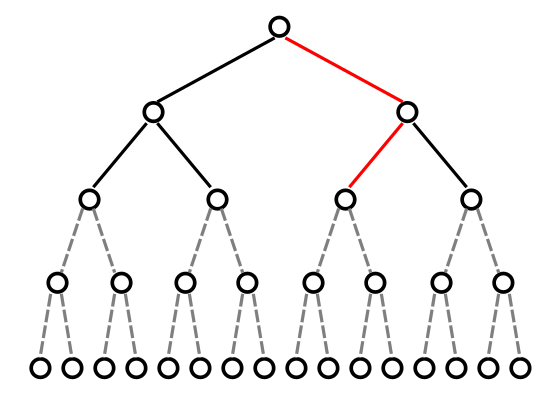}
     }
     \hfill
     \subfloat[local planning with value function\label{fig:planning_with_value_function}]{%
       \includegraphics[width=0.23\textwidth]{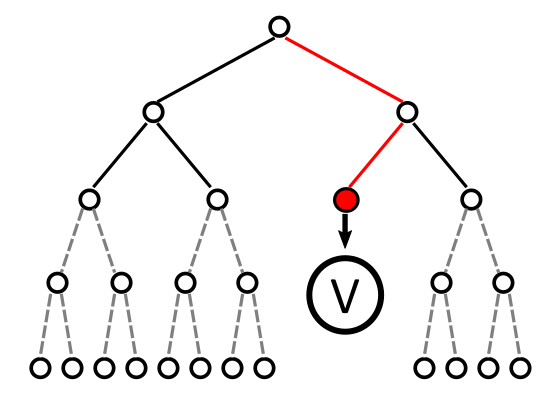}
     }
     \caption{Illustration of local planning with a horizon of $h = 2$. The nodes in the search tree represent states and the links represent actions. The red path represents a sampled plan. The dashed gray links mark unreachable paths. (a) plain local planning. (b) local planning with a value function to consider global effects in the unreachable subtree.}
     \label{fig:local_planning}
\end{figure}

\emph{Planning} searches for a policy, given a generative model $\hat{M}$, which represents the actual environment $M$. $\hat{M}$ provides an approximation for $\mathcal{P}$ and $\mathcal{R}$ of the underlying MDP or MMDP \cite{boutilier1996planning,weinstein2013open,belzner2015onplan}. We assume that $\hat{M}$ perfectly models the environment such that $\hat{M} = M$. \emph{Global planning} methods search the whole state space to find $\pi^{*}$ or $V^{*}$. An example is \emph{value iteration}, which computes the optimal value function $V^{*}$ by iteratively updating value estimates for each state according to eq. \ref{eq:policy_value_function} \cite{bellman1957,beranek1961ronald,boutilier1996planning}. \emph{Local planning} methods only regard the current state and possible future states within a horizon of $h$ to find a local policy $\pi_{\textit{local}}$ \cite{weinstein2013open,belzner2015onplan}.
An example for local planning is given in fig. \ref{fig:open_loop_planning} for a problem with a branching factor of two and a planning horizon of $h = 2$. The nodes in the search tree represent states and the links represent actions.

In this paper, we only focus on local planning methods for \emph{online planning}, where planning and execution of actions are performed alternately at each time step, given a fixed computation budget $n_{\textit{budget}}$ \cite{silver2010monte,amato2015scalable,belzner2015onplan,silver2016mastering,claes2017decentralised}.

Local planning can be performed via \emph{closed-loop} or \emph{open-loop} search. Closed-loop search corresponds to a tree search, where a search tree is constructed and traversed guided by an action selection strategy $\pi_{\textit{tree}}$ \cite{perez2015open,belzner2017stacked}. The nodes of the tree represent states and the links represent actions. The state values $V^{\pi_{\textit{tree}}}(s_{t})$ are computed recursively according to eq. \ref{eq:return} starting from the leaves of the search tree. \emph{Monte Carlo Tree Search (MCTS)} is a popular closed-loop planning approach, which is applied to very large and complex domains \cite{chaslot2010monte,kocsis2006bandit,silver2010monte,silver2016mastering,silver2017mastering}. MCTS can also be adapted to multi-agent planning \cite{amato2015scalable,claes2017decentralised}. Open-loop planning searches for action sequences or plans of length $h$ \cite{bubeck2010open,weinstein2013open,perez2015open,belzner2017stacked}. These plans are typically sampled from a sequence of distributions $\Phi_{1},...,\Phi_{h}$ and simulated in $\hat{M}$. The resulting rewards are accumulated according to eq. \ref{eq:return} and used to update the distributions. Open-loop planning does not store any information about intermediate states, thus enabling efficient planning in large-scale domains \cite{weinstein2013open,perez2015open}. An approach to open-loop planning in MAS is proposed in \cite{belzner2017scalable}.

\subsection{Reinforcement Learning}\label{subsec:reinforcement_learning}
\emph{Reinforcement Learning (RL)} corresponds to a policy search for an unknown environment $M$. In general, an agent knows the state and action space $\mathcal{S}$ and $\mathcal{A}$ but it does not know the effect of executing $a_{t} \in \mathcal{A}$ in $s_{t} \in \mathcal{S}$ \cite{boutilier1996planning,sutton1998introduction}. \emph{Model-based} RL methods learn a model $\hat{M} \approx M$ by approximating $\mathcal{P}$ and $\mathcal{R}$ \cite{boutilier1996planning,sutton1998introduction,hester2013texplore}. $\hat{M}$ can be used for planning to find a policy. In this paper, we focus on \emph{model-free} RL to approximate $V^{*}$ based on experience samples $e_{t} = (s_{t}, a_{t}, s_{t+1},r_{t})$ and a parametrized function approximator $\hat{V}_{\theta}$ with parameters $\theta$ without learning a model $\hat{M}$ \cite{sutton1998introduction}. A policy $\hat{\pi}$ can be derived by maximizing $\hat{V}_{\theta}$ such that $\hat{\pi}(s_{t}) = argmax_{a_{t} \in \mathcal{A}}(\hat{Q}_{\theta}(s_{t},a_{t}))$, where $\hat{Q}_{\theta}(s_{t},a_{t}) = \mathcal{R}(s_{t},a_{t}) + \gamma\sum_{s_{t+1} \in \mathcal{S}} \mathcal{P}(s_{t+1}|s_{t}, a_{t})\hat{V}_{\theta}(s_{t+1})$ is the approximated \emph{action value function} \cite{boutilier1996planning,sutton1998introduction}. The experience samples are obtained from interaction between the agent and the environment.

\section{Related Work}\label{sec:related_work}
\paragraph{Hybrid Planning}
Some hybrid approaches to combine offline and online planning in partially observable domains were introduced in \cite{paquet2006hybrid,ross2007aems}. In the offline planning phase, a value function $V_{MDP}$\footnote[1]{The action value function $Q(s_{t},a_{t})$ is often used instead of the state value function $V(s_{t})$. We limit our scope to the computation of $V(s_{t})$, however. } is computed based on a fully observable model of the actual environment by using variants of value iteration. $V_{MDP}$ is used to enhance online planning to search for a policy under the consideration of possible global effects. It was shown that $V_{MDP}$ provides an upper bound to $V^{*}$ of the actual environment \cite{cassandra2016learning,oliehoek2008optimal}.

This can be exploited to prune the search space without loosing optimality of the solutions found. Many multi-agent planning algorithms use similar methods to enhance planning with such a pre-computed value function $V_{MDP}$ \cite{emery2004approximate,szer2005maa,oliehoek2008optimal,spaan2011scaling}.

In our approach, $V^{*}$ is approximated \emph{online} based on \emph{actual} experience without requiring a model. A generative model is only used for online planning to find a joint policy. We intend to apply our approach to highly complex and stochastic domains, where an offline computation is not feasible, since any change in the model would require the re-computation of $V_{MDP}$.

\paragraph{Online Planning and Deep RL}
\emph{AlphaGo} is a program introduced in \cite{silver2016mastering}, which is able to play Go at a super-human level. It recently defeated the currently best human Go players in various tournaments \cite{silver2016mastering,silver2017mastering}. AlphaGo uses MCTS for online planning and deep neural networks, which approximate $\pi^{*}$ and $V^{*}$ to guide the tree search. With this approach, AlphaGo is able to develop extremely complex strategies within given time constraints.

MCTS-based planning combined with an approximation of $V^{*}$ was shown to improve the performance of plain online planning or RL in complex and deterministic games like Go and Hex \cite{silver2016mastering,silver2017mastering,thinkingfastandslow}. The idea of these approaches is based on the human mind, which is able to think ahead into the future, while guiding the thoughts with intuition learned from experience. In the context of artificial intelligence, online planning represents the future thinking, while deep RL represents the integration of strong intuition \cite{evans1984heuristic,kahneman2003maps,thinkingfastandslow}.

Our approach extends these ideas to environments with \emph{multiple} agents. We also focus on \emph{stochastic} domains, where the outcome of actions and the behaviour of agents are not deterministic.

\paragraph{Distributed Value Function Approximation}
In this paper, we focus on centralized learning of $V^{*}$, where all agents share the same parameters $\theta$ similarly to \cite{foerster2016learning,tan1997multi}. Unlike previous work on multi-agent RL, we do not use the approximated value function to directly derive a policy. Instead, we use it to guide online planning in MAS.

Besides, there exist approaches to approximate the value function asynchronously and in parallel \cite{nair2015massively,mnih2016asynchronous}. In that case, multiple agents act independently of each other in different instances of the same domain. They share experience with each other in order to update the same value function approximation $\hat{V}_{\theta}$ in parallel to accelerate the learning process.

Our approach approximates $V^{*}$ based on the global experience of multiple agents, which act in the \emph{same} environment. Our approximation $\hat{V}_{\theta}$ is not meant to improve the performance of individual agents but to improve the behaviour of the MAS as a \emph{whole}.

\section{EVADE}\label{sec:evade}

We now describe \emph{Emergent Value function Approximation for Distributed Environments (EVADE)} for leveraging statistical multi-agent online planning with a value function, which is approximated online at system runtime. EVADE is a framework for combining multi-agent online planning and RL to further improve the performance in MAS.

\subsection{Combining Online Planning and RL}
Given a perfect generative model $\hat{M} = M$, online planning can be used for decision making with high quality and accuracy w.r.t. the expected return. However, due to computational constraints, online planning is unable to make lookaheads for arbitrarily long horizons, which would be required for highly complex tasks that require much more time steps to solve than the actually feasible horizon as sketched in fig. \ref{fig:open_loop_planning}. In contrast, model-free RL with a parametrized function approximator $\hat{V}_{\theta}$ allows for potentially infinite future prediction but has approximation erros due to the compressing nature of $\hat{V}_{\theta}$.

By combining online planning and RL, a decision maker can benefit from both advantages \cite{silver2016mastering,silver2017mastering,thinkingfastandslow}. The limited lookahead of planning can be enhanced with $\hat{V}_{\theta}$ as shown in fig. \ref{fig:planning_with_value_function}. Online planning can plan accurately for $h$ initial time steps, which are weighted more than the outcome estimate $\hat{V}_{\theta}(s_{t+h})$, given a discount factor of $\gamma < 1$. The discount can also neglect possible approximation errors of $\hat{V}_{\theta}$. Especially in highly complex and stochastic domains with multiple agents, we believe that the integration of a value function approximation could improve the performance of otherwise limited multi-agent online planning.

\subsection{Multi-Agent Planning with Experience}
We focus on online settings, where there is an alternating \emph{planning} and \emph{learning} step for each time step $t$. In the planning step, the system searches for a joint policy $\pi_{\textit{local}}$, which maximizes $G_{t,\textit{EVADE}}$:
\begin{equation}\label{eq:return_local_evade}
G_{t,\textit{EVADE}} = G_{t} + \gamma^{h}\hat{V}_{\theta}(s_{t+h})
\end{equation}
$G_{t,\textit{EVADE}}$ extends $G_{t}$ from eq. \ref{eq:return} with $\hat{V}_{\theta}(s_{t+h})$ as the provided global outcome estimate to enhance local planning with a limited horizon of $h$ as sketched in fig. \ref{fig:planning_with_value_function}. The planning step can be implemented with an arbitrary multi-agent planning algorithm, depending on the concrete problem.

After the planning step, all agents execute the joint action $a_{t} = \pi_{\textit{local}}(s_{t})$ causing a state transition from $s_{t}$ to $s_{t+1}$ with a reward signal $r_{t}$. This emergent result is stored as experience sample $e_{t} = (s_{t},a_{t},s_{t+1},r_{t})$ in an experience buffer $D$. A sequence of experience samples $e_{1},...,e_{T}$ is called \emph{episode} of length $T$.

In the subsequent learning step, a parametrized function approximator $\hat{V}_{\theta}$ is used to minimize the one-step temporal difference (TD) error of all samples $e_{t}$ in $D$ w.r.t. $\theta$. The TD error for $e_{t}$ is defined by \cite{sutton1988learning,sutton1998introduction}:
\begin{equation}\label{eq:TDError}
\delta_{t} =  \hat{V}_{\theta}(s_{t}) - (r_{t} + \gamma \hat{V}_{\theta}(s_{t+1}))
\end{equation}
It should be noted that the approximation only depends on the experience samples $e_{t} \in D$ and does not require a model like hybrid planning approaches explained in section \ref{sec:related_work}. The updated value function $\hat{V}_{\theta}$ can then be used for the next planning step at $t+1$.

The complete formulation of multi-agent online planning with EVADE is given in algorithm \ref{algorithm:EVADE}, where $T$ is the length of an episode, $\hat{M}$ is the generative model used for planning, $n$ is the number of agents in the MAS, $h$ is the planning horizon, $n_{\textit{budget}}$ is the computation budget and $\hat{V}_{\theta}$ is the value function approximator. The parameter $MASPlan$ can be an arbitrary multi-agent planning algorithm for searching a joint policy $\pi_{\textit{local}}$ by maximizing $G_{t,\textit{EVADE}}$. Given that the computation budget $n_{\textit{budget}}$ is fixed and the time to update $\hat{V}_{\theta}$ at each time step is constant\footnote[2]{In practice, $\theta$ is updated w.r.t. experience batches of constant size, which are sampled from $D$ \cite{mnih2013playing,mnih2015human}.}, EVADE is suitable for online planning and learning in real-time MAS.

\begin{algorithm}
\caption{Multi-agent online planning with EVADE}\label{algorithm:EVADE}
\begin{algorithmic}[1]
\Procedure{$EVADE(MASPlan, \hat{M}, n, h, n_{\textit{budget}}, \hat{V}_{\theta})$}{}
\State Initialize $\theta$ of $\hat{V}_{\theta}$
\State Observe $s_{1}$
\For{$t = 1,T$}
\State	Find $\pi_{\textit{local}}$ using $MASPlan(s_{t}, \hat{M}, n, h, n_{\textit{budget}}, \hat{V}_{\theta})$
\State	Execute $a_{t} = \pi_{\textit{local}}(s_{t})$
\State  Observe reward $r_{t}$ and new state $s_{t+1}$
\State	Store new experience $e_{t} = (s_{t}, a_{t}, s_{t+1}, r_{t})$ in $D$
\State	Refine $\theta$ to minimize the TD error $\delta_{t}$ for all $e_{t} \in D$
\EndFor
\EndProcedure
\end{algorithmic}
\end{algorithm}

\subsection{Architecture}\label{subsec:EVADE_architecture}

\begin{figure}[!ht]
     \subfloat[Centralized planning\label{fig:evade_centralized_planning}]{%
       \includegraphics[width=0.225\textwidth]{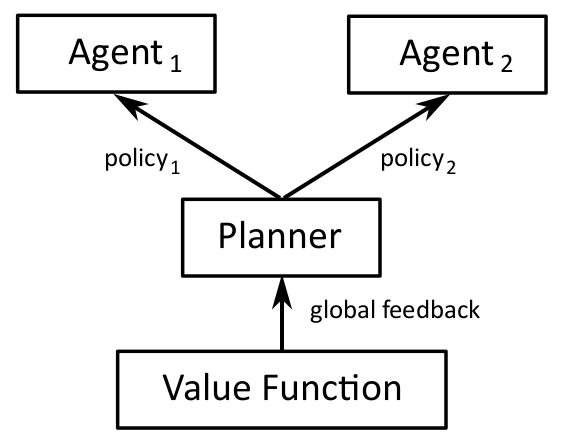}
     }
     \hfill
     \subfloat[Decentralized planning\label{fig:evade_decentralized_planning}]{%
       \includegraphics[width=0.225\textwidth]{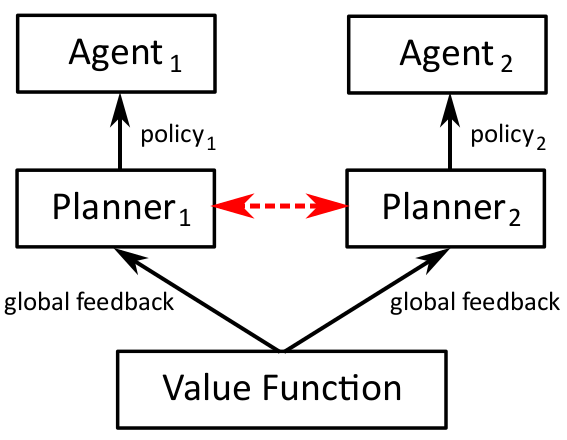}
     }
     \caption{Illustration of the possible MAS planning architectures for EVADE. The planners get global feedback from a value function, which is approximated in a centralized way. The red dashed arrow between the planners in fig. \ref{fig:evade_decentralized_planning} represents a coordination mechanism for decentralized planning.}
     \label{fig:evade_architecture}
\end{figure}

We focus on centralized learning, since we believe that $V^{*}$ can be approximated faster if all agents share the same parameters $\theta$ \cite{tan1997multi,foerster2016learning}. Online planning can be performed in a centralized or decentralized way by using a concrete MAS planning algorithm. In both cases, each planner uses the common value function approximation $\hat{V}_{\theta}$ to search for $\pi_{\textit{local}}$ by maximizing $G_{t,\textit{EVADE}}$. A conceptual overview of the EVADE architecture is shown in fig. \ref{fig:evade_architecture}. Completely decentralized architectures, where all agents plan and learn independently of each other, are not considered here and left for future work.

Decentralized planning approaches require an explicit coordination mechanism to avoid convergence to suboptimal joint policies as shown in fig. \ref{fig:evade_decentralized_planning} and in \cite{boutilier1996planning,bucsoniu2010multi}. This could be done by using a consensus mechanism to \emph{synchronize} on time or on a common seed value to generate the same random numbers when sampling plans \cite{emery2004approximate}. Agents could also exchange observations, experience, plans or policies via \emph{communication} \cite{tan1997multi,wu2009multi}. Another way is to \emph{predict} other agents' actions by using a policy function similarly to \cite{silver2016mastering} or by maintaining a belief about other agents' behaviour \cite{bucsoniu2010multi,oliehoek2016concise}.

\section{Experiments}\label{sec:experiments}
\subsection{Evaluation Environment}\label{subsec:environment}
\subsubsection{Description}
\begin{figure}[!ht]
     \subfloat[machine grid\label{fig:machine_grid}]{%
       \includegraphics[width=0.2\textwidth]{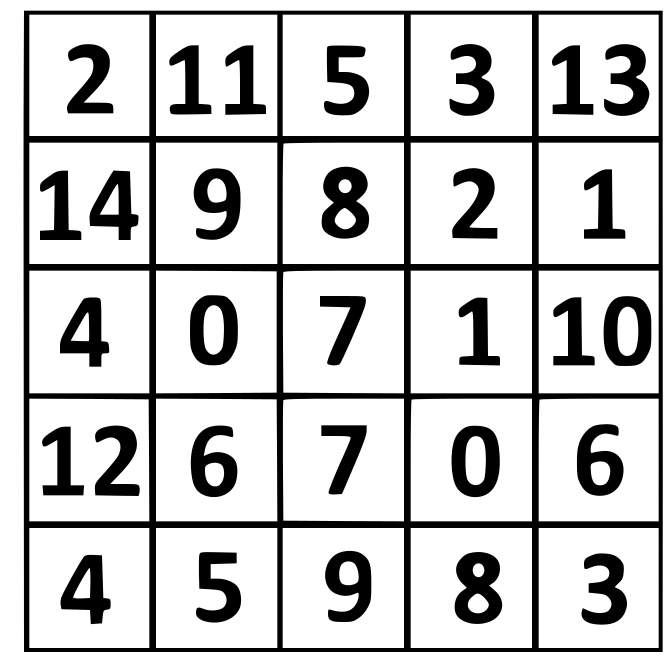}
     }
     \hfill
     \subfloat[an agent and its tasks\label{fig:item_tasks_example}]{%
       \includegraphics[width=0.2\textwidth]{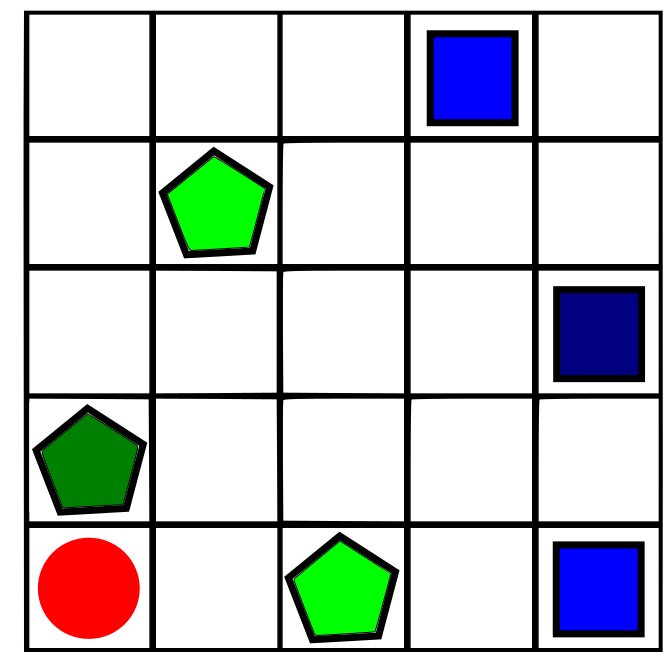}
     }
     \caption{Illustration of the smart factory setup used in the experiments. (a) the $5 \times 5$ grid of machines. The numbers in each grid cell denote the machine type. (b) an agent $i$ (red circle) in the factory with $tasks_{i} = [\{9,12\},\{3,10\}]$. It should get processed at the green pentagonal machines first before going to the blue rectangular machines.}
     \label{fig:factory_setup}
\end{figure}

We implemented a smart factory environment to evaluate multi-agent online planning with EVADE. Our smart factory consists of a $5 \times 5$ grid of \emph{machines} with 15 different machine types as shown in fig. \ref{fig:machine_grid}. Each \emph{item} is carried by one agent $i$ and needs to get processed at various machines according to its randomly assigned processing tasks $tasks_{i} = [\{a_{i,1},b_{i,1}\}, \{a_{i,2},b_{i,2}\}]$, where each task $a_{i,j}$,$b_{i,j}$ is contained in a \emph{bucket}. While tasks in the same bucket can be processed in any order, buckets themselves have to be processed in a specific order. Fig. \ref{fig:item_tasks_example} shows an example for an agent $i$ with $tasks_{i} = [\{9,12\},\{3,10\}]$. It first needs to get processed by the machines marked as green pentagons before going to the machines marked as blue rectangles. Note that $i$ can choose between two different machines for processing its requests $a_{i,1} = 9$ and $a_{i,2} = 3$, which are rendered as light green pentagons or light blue rectangles. In the presence of multiple agents, coordination is required to choose an appropriate machine, while avoiding conflicts with other agents.

All agents have a random initial position and can move along the machine grid or enqueue at their current position represented by a machine. Each machine can process exactly one item per time step with a cost of 0.25 but fails with a probability of 0.1 to do so. Enqueued agents are unable to perform any actions. If a task is processed, it is removed from its bucket. If a bucket is empty, it is removed from the item's tasks list. An item is \emph{complete} if its tasks list is empty. The goal is to complete as many items as possible within 50 time steps, while avoiding any conflicts or enqueuing at wrong machines.

\subsubsection{MMDP Formulation}

The smart factory environment can be modeled as MMDP $M = \langle\mathcal{D},\mathcal{S},\mathcal{A},\mathcal{P},\mathcal{R}\rangle$.
$\mathcal{D}$ is the set of $n$ agents with $\mathcal{D}_{\textit{active}} \cap \mathcal{D}_{\textit{complete}} = \emptyset$ and $\mathcal{D} = \mathcal{D}_{\textit{active}} \cup \mathcal{D}_{\textit{complete}}$. $\mathcal{D}_{\textit{active}}$ is the set of agents with incomplete items and $\mathcal{D}_{\textit{complete}}$ is the set of agents with complete items.
$\mathcal{S}$ is a set of system states described by the individual state variables of all agents, items and machines.
$\mathcal{A}$ is the set of joint actions. Each agent $i \in \mathcal{D}$ has the same individual action space $\mathcal{A}_{i}$ enabling it to move north, south, west or east, to enqueue at its current machine $m = pos_{i}$ or to do nothing. Any attempt to move across the grid boundaries is treated the same as "do nothing".
$\mathcal{P}$ is the transition probability function.
$\mathcal{R}$ is the scalar reward function. $\mathcal{R}$ at time step $t$ is defined by $R(s_{t},a_{t}) = score_{t+1} - score_{t}$, where $score_{t}$ is the immediate evaluation function for the system state:
\begin{equation}\label{eq:score_function}
score_{t} = |\mathcal{D}_{\textit{complete}}| - tasks_{t} - cost_{t} - tpen_{t}
\end{equation}
where $tasks_{t} = \sum_{i \in \mathcal{D}_{\textit{active}}}^{}\sum_{c \in tasks_{i}}^{} |c|$ is the total number of currently unprocessed tasks, $cost_{t}$ is the total sum of processing costs for each machine after processing an enqueued item and $tpen_{t} = tpen_{t-1} + \sum_{i \in \mathcal{D}_{\textit{active}}}^{} penalty$ is the total sum of time penalities with $penalty = 0.1$ for all incomplete items at time step $t$. Processing tasks and completing items increases $score_{t}$. Otherwise, $score_{t}$ decreases for each incomplete item or enqueuing at a wrong machine.

\subsubsection{Complexity}

Depending on the number of agents $n$, the number of possible joint actions is $|\mathcal{A}| = |\mathcal{A}_{i}|^{n} = 6^{n}$. The machine failure probability of 0.1 increases the branching factor of the problem even more. Given a planning horizon of $h$, the number of possible joint plans is defined by:
\begin{equation}\label{eq:factory_search_space}
|\pi_{\textit{local}}| = |\Pi_{local,i}|^{n} = (|\mathcal{A}_{i}|^{h})^{n} = |\mathcal{A}_{i}|^{h \cdot n} = 6^{h \cdot n}
\end{equation}

We tested EVADE in settings with 4 and 8 agents. In the 4-agent case, there exist $6^4 \approx 1300$ possible joint actions. In the 8-agent case, there exist $6^8 \approx 1.68 \cdot 10^{6}$ possible joint actions. In our stochastic smart factory setup, where machines can fail with a probability of 0.1 and where agents are not acting in a deterministic way, the environment has a significantly higher branching factor than the game of Go, which has a branching factor of 250 \cite{silver2016mastering}.

\subsection{Methods}

\subsubsection{Online Open-Loop Planning}\label{subsubsec:online_open_loop_planning}
Due to the stochasticity and high complexity of our environment, we focus on open-loop planning because we think that current state-of-the-art algorithms based on closed-loop planning would not scale very well in our case \cite{perez2015open,amato2015scalable}. Also, we do not aim for optimal planning, since our goal is to enhance existing local planning algorithms, which might even perform suboptimal in the first place.

The individual policy $\pi_{i}$ for each agent $i$ is implemented as a stack or sequence of multi-armed bandits (MAB) of length $h$ as proposed in \cite{belzner2017stacked}. Each MAB $\Phi_{t} = P(a_{t}|D_{a_{t}})$ represents a distribution, where $D_{a_{t}}$ is a buffer of size 10 for storing local returns, which are observed when selecting arm $a_{t} \in \mathcal{A}$. Each buffer $D_{a_{t}}$ is implemented in a sliding window fashion to consider only most recent observations to adapt to the non-stationary joint behaviour of all agents during the planning step.

Thompson Sampling is implemented as concrete MAB algorithm because of its effectiveness and robustness for making decisions under uncertainty \cite{thompson1933likelihood,chapelle2011empirical,belzner2017stacked}. The implementation is adopted from \cite{honda2014optimality,bai2014thompson}, where the return values in $D_{a_{t}}$ for each arm $a_{t}$ are assumed to be normally distributed.

To optimize $\pi_{i}$, a plan of $h$ actions is sampled from the MAB stack. The plan is evaluated in a simulation by using a generative model $\hat{M}$. The resulting rewards are accumulated to local returns according to eq. \ref{eq:return_local_evade} and used to update the corresponding MABs of the MAB stack. This procedure is repeated \(\lfloor\frac{n_{\textit{budget}}}{h}\rfloor\) times. Afterwards, the action $a_{t} = argmax_{a_{1} \in \mathcal{A}}\{\overline{D_{a_{1}}}\}$ is selected from the MAB $\Phi_{1}$ for execution in the actual environment, where $\overline{D_{a_{1}}}$ is the mean of all local returns currently stored in $D_{a_{1}}$.

\subsubsection{Multi-Agent Planning}\label{subsubsec:experiments_map_algorithms}
We implemented two multi-agent planning algorithms to evaluate the performance achieved by using EVADE. All algorithms enhanced with EVADE were compared with their non-enhanced counterparts w.r.t. performance and efficiency.

\paragraph{Direct Cross Entropy (DICE) method for policy search in distributed models}
DICE is a centralized planning algorithm proposed in \cite{oliehoek2008cross} and uses stochastic optimization to search joint policies, which are optimal or close to optimal. In DICE a multivariate distribution $f_{\xi}(\pi) = \prod_{i=1}^{n}f_{\xi_{i}}(\pi_{i})$ is maintained to sample candidate joint policies \(\pi\). These candidates are evaluated in a simulation with a global model $\hat{M}$. The $N_{b}$ best candidates are used to update $f_{\xi}$. This procedure is repeated until convergence is reached or $n_{\textit{budget}}$ has run out. Our implementation of DICE uses $n$ MAB stacks representing $f_{\xi}(\pi)$ to sample joint plans of length $h$, which are simulated in $\hat{M}$. The resulting local returns are used to update all MAB stacks.

\paragraph{Distributed Online Open-Loop Planning (DOOLP)}
DOOLP is a decentralized version of DICE proposed in \cite{belzner2017scalable}, where each agent is controlled by an individual planner with an individual model $\hat{M}_{i} = \hat{M}$ for simulation-based planning. At every time step each agent $i$ iteratively optimizes its policy $\pi_{i}$ by first sampling a plan and then querying the sampled plans of its neighbours to construct a joint plan. The joint plan is simulated in $\hat{M}_{i}$ and the simulation result is used to update the individual policy $\pi_{i}$ of agent $i$.
The individual MAB stacks are assumed to be \emph{private} for each agent $i$. Due to the stochasticity of the environment described in section \ref{sec:introduction} and \ref{subsec:environment}, the planners can have different simulation outcomes leading to different updates to the individual MAB stacks.

As a decentralized approach, DOOLP requires an explicit coordination mechanism to avoid suboptimal joint policies (see section \ref{subsec:EVADE_architecture} and fig. \ref{fig:evade_decentralized_planning}). We implemented a communication-based coordination mechanism, where each planner communicates its sampled plans to all other planners, while keeping its actual MAB stack private.

\subsubsection{Value Function Approximation}
We used a deep convolutional neural network as $\hat{V}_{\theta}$ to approximate the value function $V^{*}$. The weights of the neural network are denoted as $\theta$. $\hat{V}_{\theta}$ was trained with TD learning by using methods of deep RL \cite{mnih2013playing,mnih2015human}. An experience buffer $D$ was implemented to uniformly sample minibatches to perform stochastic gradient descent on. $D$ was initialized with 5000 experience samples generated from running smart factory episodes using multi-agent planning without EVADE.

An additional target network $\hat{V}_{\theta^{-}}$ was used to generate TD regression targets for $\hat{V}_{\theta}$ (see eq. \ref{eq:TDError}) to stabilize the training \cite{mnih2015human}. All hyperparameters used for training $\hat{V}_{\theta}$ are listed in table \ref{tab:value_network_hyperparameter}.

\begin{center}
\begin{table}[htbp]
{\small
\begin{center}
\begin{tabular}[center]{|l|P{1.5cm}|} \hline
hyperparameter & value \\ \hline
update rule for optimization & ADAM \\
learning rate & 0.001 \\
discount factor \(\gamma\) & 0.95 \\
minibatch size & 64\\
replay memory size & 10000 \\
target network update frequency \(C\) & 5000 \\ \hline
\end{tabular}
\end{center}
} % end of tiny
\caption[Hyperparameters for the value network.]{Hyperparameters for the value network \(\hat{V}_{\theta}\).\label{tab:value_network_hyperparameter}}
\end{table}
\end{center}

The factory state is encoded as a stack of $5 \times 5$ feature planes, where each plane represents the spatial distribution of machines or agents w.r.t. some aspect. An informal description of all feature planes is given in table \ref{tab:feature_plane_stack}.

\begin{table*}
\centering
\caption{Description of all feature planes as input for \(\hat{V}_{\theta}\).}
\begin{tabular}[center]{|p{2.70cm}|P{1.30cm}|p{11cm}|} \hline
Feature & \# Planes & Description \\ \hline
Machine type & 1 & The type of each machine as a value between 0 and 14 (see fig. \ref{fig:machine_grid}) \\
Agent state & 4 & The number of agents standing at machines whose types are (not) contained in their current tasks and whether they are enqueued or not. \\
Tasks (1st bucket) & 15 & Spatial distribution of agents containing a particular machine type in their first bucket of tasks for each available machine type. \\
Tasks (2nd bucket) & 15 & Same as "Tasks (1st bucket)" but for the second bucket of tasks. \\ \hline
\end{tabular}\label{tab:feature_plane_stack}
\end{table*}

The input to $\hat{V}_{\theta}$ is a $5 \times 5 \times 35$ matrix stack consisting of 35 matrices. The first hidden layer convolves 128 filters of size $5 \times 5$ with stride 1. The next three hidden layer convolve 128 filters of size $3 \times 3$ with stride 1. The fifth hidden layer convolves one filter of size $1 \times 1$ with stride 1. The sixth hidden layer is a fully connected layer with 256 units. The output layer is a fully connected with a single linear unit. All hidden layers use exponential linear unit (ELU) activation as proposed in \cite{clevert2015fast}. The architecture of $\hat{V}_{\theta}$ was inspired by the value network of \cite{silver2016mastering}. 

\subsection{Results}
Various experiments with 4- and 8-agent settings were conducted to study the effectiveness and efficiency achieved by the multi-agent online planning algorithms from section \ref{subsubsec:experiments_map_algorithms} with EVADE.

An episode is reset after $T = 50$ time steps or when all items are complete such that $\mathcal{D}_{\textit{active}} = \emptyset$. A run consists of 300 episodes and is repeated 100 times.
Multi-agent online planning with EVADE searches for a joint policy $\pi_{\textit{local}}$ by maximizing $G_{t,\textit{EVADE}}$ with a value function approximation $\hat{V}_{\theta}$ (see eq. \ref{eq:return_local_evade}). All baselines perform planning without EVADE by maximizing $G_{t}$ instead (see eq. \ref{eq:return}).

The performance of multi-agent online planning is evaluated with the value of $score_{50}$ at the end of each episode (see eq. \ref{eq:score_function}) and the item completion rate $R_{\textit{completion}}$ at the end of the $300^{th}$ episode,  which is defined by:
\begin{equation}
R_{\textit{completion}} = \frac{|\mathcal{D}_{\textit{complete}}|}{|\mathcal{D}|} = \frac{|\mathcal{D}_{\textit{complete}}|}{|\mathcal{D}_{\textit{complete}} \cup \mathcal{D}_{\textit{active}}|}
\end{equation}
with $0 \leq R_{\textit{completion}} \leq 1$. If all items are complete within 50 time steps, then $R_{\textit{completion}} = 1$. If no item is complete within 50 time steps, then $R_{\textit{completion}} = 0$. All baselines were run 500 times to determine the average of $score_{50}$ and $R_{\textit{completion}}$.

\subsubsection{Efficiency w.r.t. Computation Budget}
The effect of EVADE w.r.t. the breadth of the policy search was evaluated. The experiments for each algorithm were run with different budgets $n_{\textit{budget}} \in \{192, 384, 512\}$\footnote[3]{We also experimented with $n_{\textit{budget}} = 256$ but there was no significant difference to planning with $n_{\textit{budget}} = 384$.} and a fixed horizon of $h = 4$. The baselines represented by the corresponding non-enhanced planning algorithms had a computation budget of $n_{\textit{budget}} = 512$.

\begin{figure}[!ht]
     \subfloat[DICE (4 agents)\label{fig:learning_progress_dice_4_agents}]{%
       \includegraphics[width=0.23\textwidth]{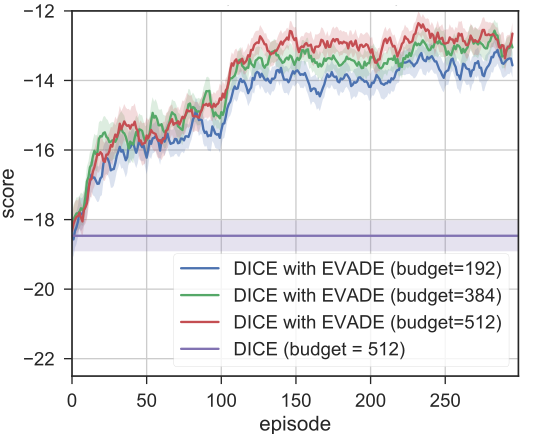}
     }
     \hfill
     \subfloat[DOOLP (4 agents)\label{fig:learning_progress_doolp_4_agents}]{%
       \includegraphics[width=0.23\textwidth]{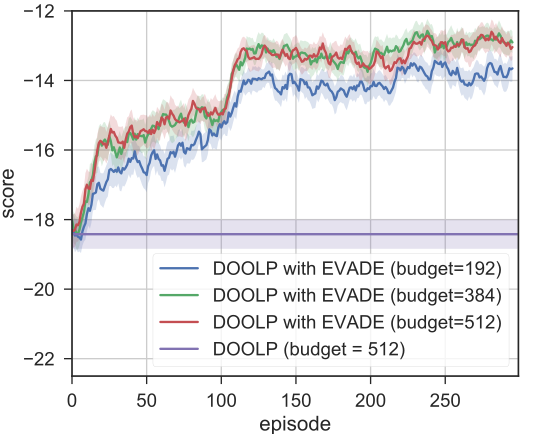}
     }
     \\
     \subfloat[DICE (8 agents)\label{fig:learning_progress_dice_8_agents}]{%
       \includegraphics[width=0.23\textwidth]{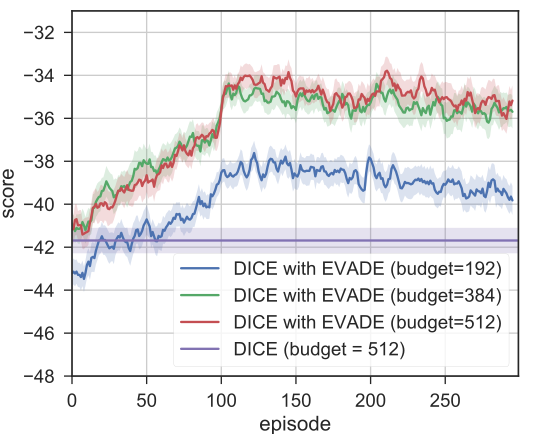}
     }
     \hfill
     \subfloat[DOOLP (8 agents)\label{fig:learning_progress_doolp_8_agents}]{%
       \includegraphics[width=0.23\textwidth]{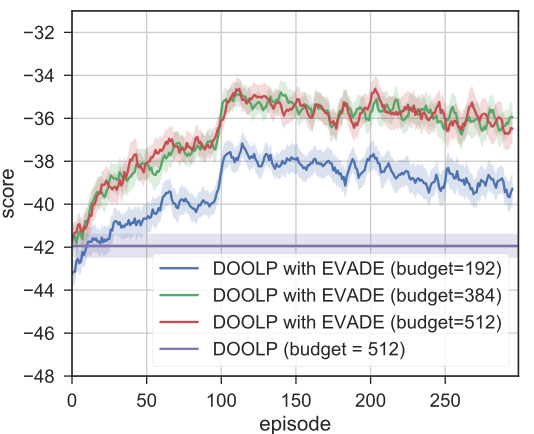}
     }
     \caption{Average progress of $score_{50}$ of 100 runs shown as running mean over 5 episodes for different computation budgets $n_{\textit{budget}} \in \{192, 384, 512\}$ and a horizon of $h = 4$. All baselines have a computation budget of $n_{\textit{budget}} = 512$. Shaded areas show the 95\% confidence interval.}
     \label{fig:evade_budget_efficiency}
\end{figure}

Fig. \ref{fig:evade_budget_efficiency} shows the average progress of $score_{50}$. In all cases, the EVADE enhanced versions outperform their corresponding baselines. There is a relatively large performance gain in the first hundred episodes. The average score increases slowly afterwards or stagnates as shown in the 8-agent case in fig. \ref{fig:learning_progress_dice_8_agents} and \ref{fig:learning_progress_doolp_8_agents}.
There are no significant differences between the enhanced versions with $n_{\textit{budget}} \in \{384, 512\}$. Planning with a budget of $n_{\textit{budget}} = 192$ leads to worse performance than the corresponding enhanced variants with a larger budget.

The average completion rates $R_{\textit{completion}}$ at the end of the $300^{th}$ episode of all experiments are listed in table \ref{tab:completion_rate_budget}. In the 4-agent case, the completion rates of the baselines are about 63\%, while the rates achieved by the EVADE enhanced versions range from 86 to 92\%. In the 8-agent case, the completion rates of the baslines are about 54\%, while the rates achieved by the EVADE enhanced versions range from 65 to 78\%. EVADE enhanced planning with $n_{\textit{budget}} \in \{384, 512\}$ always tends to achieve a higher completion rate than using a budget of $n_{\textit{budget}} = 192$.

\begin{table*}
\centering
\caption{Average rate of complete items $R_{\textit{completion}}$ at the end of the $300^{th}$ episode of all experiments within a 95\% confidence interval. Planning was performed with different computation budgets $n_{\textit{budget}}$ and a horizon of $h = 4$.}
\begin{tabular}{|P{4cm}|P{2.75cm}|P{2.75cm}|P{2.75cm}|P{2.75cm}|} \hline
algorithm (\# agents) & baseline ($n_{\textit{budget}}=512$) & EVADE ($n_{\textit{budget}}=192$) & EVADE ($n_{\textit{budget}}=384$) & EVADE ($n_{\textit{budget}}=512$)\\ \hline
DICE (4 agents) & $62.5 \pm 2.1\%$ & $86.8 \pm 3.6\%$ & $89.3 \pm 3.0\%$ & \bm{$91.8 \pm 3.0 \%$} \\
DOOLP (4 agents) & $63.7 \pm 2.1\%$ & $88.5 \pm 3.0\%$ & \bm{$91.3 \pm 2.9\%$} & $91.0 \pm 3.4\%$ \\ \hline
DICE (8 agents) & $55.2 \pm 1.5\%$ & $65.0 \pm 3.4\%$ & $73.1 \pm 3.3\%$ & \bm{$77.5 \pm 3.2\%$} \\
DOOLP (8 agents) & $53.9 \pm 1.4\%$ & $65.8 \pm 3.7\%$ & $72.8 \pm 3.6\%$ & \bm{$73.0 \pm 3.5\%$} \\ \hline
\end{tabular}\label{tab:completion_rate_budget}
\end{table*}

\subsubsection{Efficiency w.r.t. Horizon}\label{subsubsec:efficiency_horizon}
Next the effect of EVADE w.r.t. the depth of the policy search was evaluated. The experiments for each algorithm were run with different horizon lengths $h \in \{2, 4, 6\}$ and a fixed computation budget of $n_{\textit{budget}} = 384$. The baselines represented by the corresponding non-enhanced planning algorithms had a horizon of $h = 6$.

\begin{figure}[!ht]
     \subfloat[an agent and its tasks\label{fig:full_visibility}]{%
       \includegraphics[width=0.2\textwidth]{img/grid_map_example.png}
     }
     \hfill
     \subfloat[horizon of $h = 2$\label{fig:horizon_2}]{%
       \includegraphics[width=0.2\textwidth]{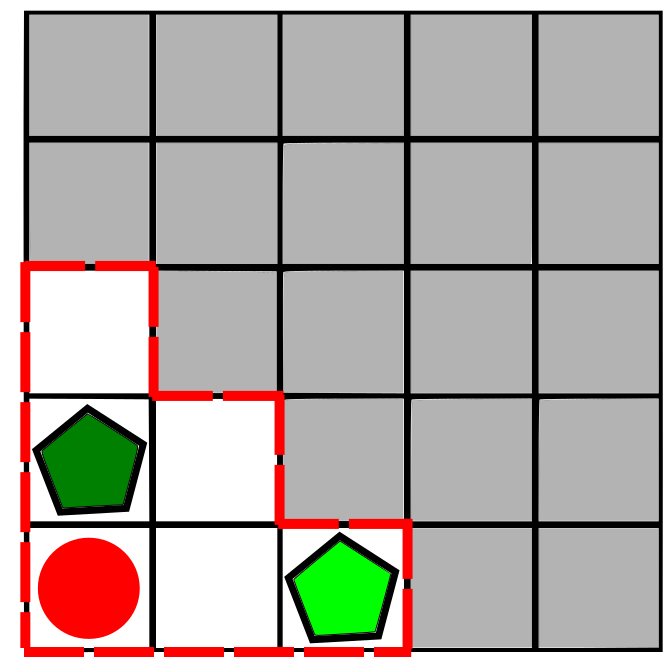}
     } \\
     \subfloat[horizon of $h = 4$\label{fig:horizon_4}]{%
       \includegraphics[width=0.2\textwidth]{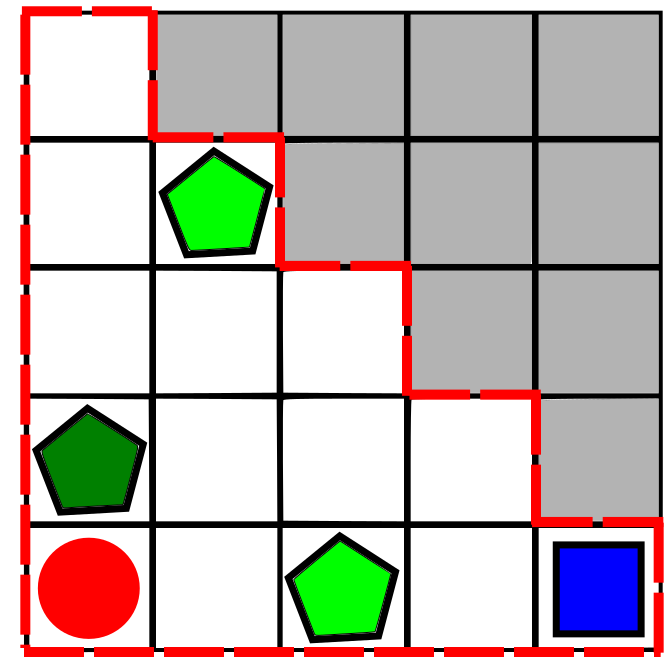}
     }
     \hfill
     \subfloat[horizon of $h = 6$\label{fig:horizon_6}]{%
       \includegraphics[width=0.2\textwidth]{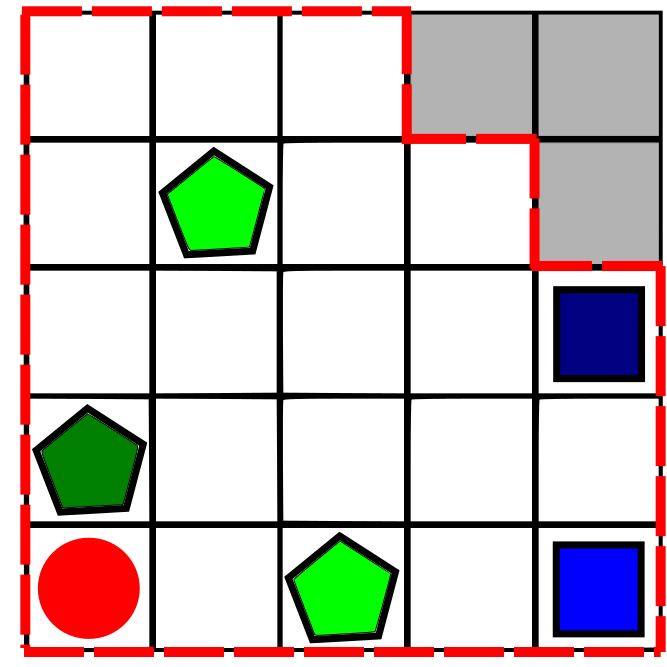}
     }
     \caption{Reachability of machines for an agent (red circle) in a simulation step depending on the planning horizon $h$. Gray grid cells mark unreachable machines. (a) The example from fig. \ref{fig:item_tasks_example}. (b), (c) and (d) Reachable machines within the dashed red boundaries, given resp. horizons of $h$.}
     \label{fig:factory_reachable_spots}
\end{figure}

The planning horizon $h$ influences the reachability of machines in each simulation step as shown in fig. \ref{fig:factory_reachable_spots}. In this example, the agent can only reach about one fifth of the grid when planning with $h = 2$ (see fig. \ref{fig:horizon_2}), while it can theoretically reach almost any machine when planning with $h = 6$ (see fig. \ref{fig:horizon_6}).

\begin{figure}[!ht]
     \subfloat[DICE (4 agents)\label{fig:learning_progress_horizon_dice_4_agents}]{%
       \includegraphics[width=0.23\textwidth]{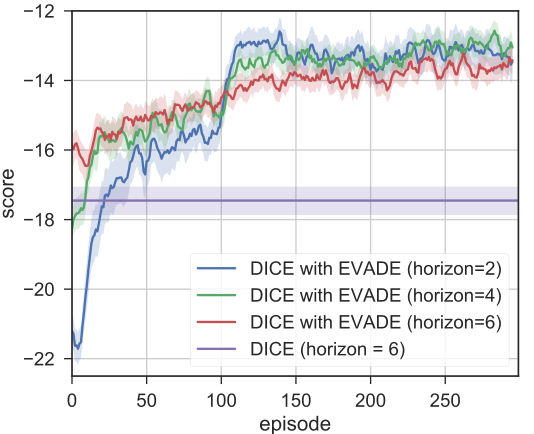}
     }
     \hfill
     \subfloat[DOOLP (4 agents)\label{fig:learning_progress_horizon_doolp_4_agents}]{%
       \includegraphics[width=0.23\textwidth]{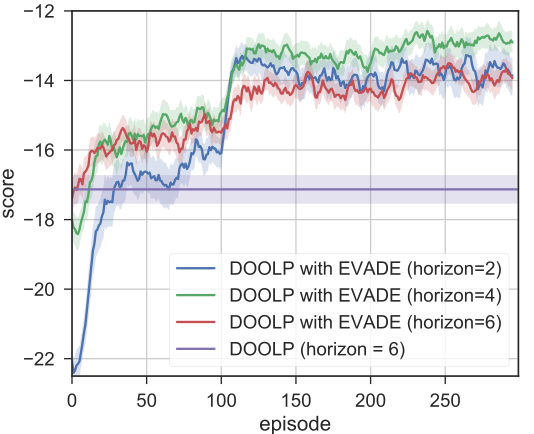}
     }
     \\
     \subfloat[DICE (8 agents)\label{fig:learning_progress_horizon_dice_8_agents}]{%
       \includegraphics[width=0.23\textwidth]{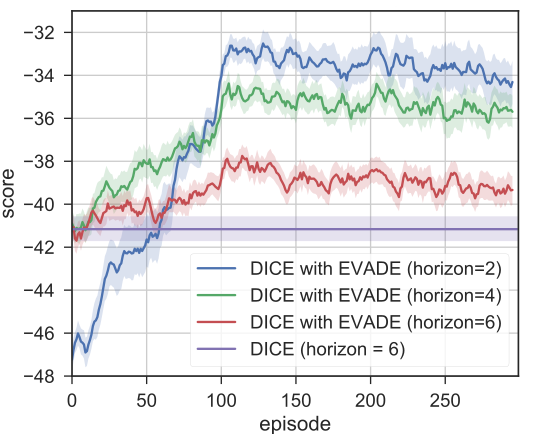}
     }
     \hfill
     \subfloat[DOOLP (8 agents)\label{fig:learning_progress_horizon_doolp_8_agents}]{%
       \includegraphics[width=0.23\textwidth]{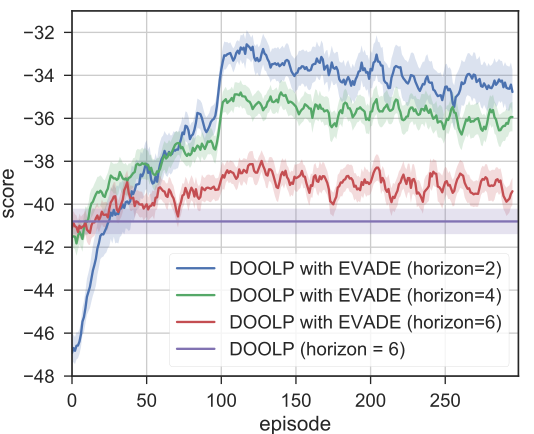}
     }
     \caption{Average progress of $score_{50}$ of 100 runs shown as running mean over 5 episodes for different horizons $h \in \{2, 4, 6\}$ and a computation budget of $n_{\textit{budget}} = 384$. All baselines have a horizon of $h = 6$. Shaded areas show the 95\% confidence interval.}
     \label{fig:evade_horizon_efficiency}
\end{figure}

Fig. \ref{fig:evade_horizon_efficiency} shows the average progress of $score_{50}$. Planning with a horizon of $h = 2$ always had the worst initial average performance but the largest performance gain in the first hundred episodes, while planning with a horizon of $h = 6$ had the best initial average performance but the smallest performance gain. In the 8-agent case, planning with EVADE and a horizon of $h = 2$ even outperforms the planning variants with a longer horizon after about one hundred episodes as shown in fig. \ref{fig:learning_progress_horizon_dice_8_agents} and \ref{fig:learning_progress_horizon_doolp_8_agents}. This phenomenon will be discussed in the next section.

The average completion rates $R_{\textit{completion}}$ at the end of the $300^{th}$ episode of all experiments are listed in table \ref{tab:completion_rate_horizon}.
In the 4-agent case, the completion rate of the baselines are about 70\%, while the rate achieved by the EVADE enhanced versions range from about 82 to 92\%. In the 8-agent case, the completion rates of the baslines are about 59\%, while the rate achieved by the EVADE enhanced versions range from about 66 to 77\%. Increasing the horizon from 2 to 6 in the 4-agent case tends to slightly increase $R_{\textit{completion}}$, while in the 8-agent case it leads to a decrease of $R_{\textit{completion}}$.

\begin{table*}
\centering
\caption{Average rate of complete items $R_{\textit{completion}}$ at the end of the $300^{th}$ episode of all experiments within a 95\% confidence interval. Planning was performed with different horizons $h$ and a computation budget of $n_{\textit{budget}} = 384$.}
\begin{tabular}{|P{4cm}|P{2.75cm}|P{2.75cm}|P{2.75cm}|P{2.75cm}|} \hline
algorithm (\# agents) & baseline ($h = 6$) & EVADE ($h = 2$) & EVADE ($h = 4$) & EVADE ($h = 6$)\\ \hline
DICE (4 agents) & $69.7 \pm 2.0\%$ & $87.0 \pm 3.2\%$ & $89.3 \pm 3.0\%$ & \bm{$90.8 \pm 3.2\%$}\\
DOOLP (4 agents) & $71.6 \pm 2.0\%$ & $82.3 \pm 4.0\%$ & \bm{$91.3 \pm 2.9\%$} & $88.5 \pm 3.4\%$\\ \hline
DICE (8 agents) & $58.3 \pm 1.5\%$ & \bm{$77.0 \pm 3.8\%$} & $73.1 \pm 3.3\%$ & $66.1 \pm 3.1\%$\\
DOOLP (8 agents) & $60.0 \pm 1.5\%$ & \bm{$72.9 \pm 3.5\%$} & $72.8 \pm 3.6\%$ & $67.6 \pm 2.9\%$\\ \hline
\end{tabular}\label{tab:completion_rate_horizon}
\end{table*}

\subsection{Discussion}
Our experiments show that statistical multi-agent online planning can be effectively improved with EVADE, even when using a smaller computation budget $n_{\textit{budget}}$ than planning without any value function. However, $n_{\textit{budget}}$ must not be too small, since statistical online planning algorithms always require a minimum of computation to reach promising states with higher probability. This is shown in the experimental settings with $n_{\textit{budget}} = 192$ in fig. \ref{fig:evade_budget_efficiency} and table \ref{tab:completion_rate_budget}.

In the smart factory environment, planning with a sufficient horizon length is crucial to find joint policies with high quality as shown in fig. \ref{fig:factory_reachable_spots} and table \ref{tab:completion_rate_budget} and \ref{tab:completion_rate_horizon} regarding the performance of the baselines. If a needed machine is unreachable in the simulation, it cannot be considered in the local planning process, thus possibly leading to poor solutions. In our experiments, the value function approximation could improve multi-agent planning with horizons which were too short to consider the entire factory.

If the discount factor is $\gamma < 1$, then the value function influences planning with short horizons more than planning with a long horizon (see eq. \ref{eq:return_local_evade}). In our experiments, planning with a horizon of $h = 2$ was able to keep up with planning variants with a longer horizon, even outperforming them in the 8-agent case, given an equal computation budget of $n_{\textit{budget}} = 384$. These are strong indications that our approach offers planning efficiency w.r.t. the breadth and the depth of the policy search after a sufficient learning phase.

The performance stagnation in the 8-agent case after hundred episodes can be explained with the enormous policy space to be searched and the relatively small computation budget $n_{\textit{budget}}$. This also explains the rather poor performance of online planning with a horizon of $h = 6$ compared to variants with $h = 2$ or $h = 4$ as shown in fig. \ref{fig:learning_progress_horizon_dice_8_agents} and \ref{fig:learning_progress_horizon_doolp_8_agents}. Given $n_{\textit{budget}} = 384$, the former only performs $\lfloor\frac{n_{\textit{budget}}}{h}\rfloor = 64$ simulations per time step, while searching a much larger policy space ($|\pi_{\textit{local}}| > 10^{37}$) than the latter ($|\pi_{\textit{local}}| < 10^{25}$) according to eq. \ref{eq:factory_search_space}. When using the value function approximation $\hat{V}_{\theta}$, more simulations should lead to high quality results with a higher accuracy. Thus, a larger performance gain can be expected when increasing $n_{\textit{budget}}$.

\section{Conclusion \& Future Work}\label{sec:conlusion}
In this paper, we presented EVADE, an approach to effectively improve the performance of statistical multi-agent online planning in stochastic domains by integrating global experience. For this purpose, a value function is approximated online based on the emergent system behaviour by using model-free RL. By considering global outcome estimates with that value function during the planning step, multi-agent online planning with EVADE is able to overcome the limitation of local planning as sketched in fig. \ref{fig:local_planning}.

We also introduced a smart factory environment, where multiple agents need to process various items with different tasks at a shared set of machines in an automated and self-organizing way. Unlike domains used in \cite{silver2016mastering,silver2017mastering,thinkingfastandslow}, our environment can have multiple agents, is stochastic % w.r.t. the outcome of actions and the behaviour of agents. It also has a higher branching factor, given a sufficient number of agents.
and has a higher branching factor, given a sufficient number of agents.

EVADE was successfully tested with two existing statistical multi-agent planning algorithms in this highly complex and stochastic domain. EVADE offers planning efficiency w.r.t. the depth and the breadth of the joint policy search requiring less computational effort to find solutions with higher quality compared to multi-agent planning without any value function.

For now, EVADE has only been applied to fully observable settings. Decentralized partially observable problems can often be decomposed into smaller subproblems, which are fully observable themselves. This is common in distributed environments, where agents can sense and communicate with all neighbours within their range. EVADE could be directly applied to those subproblems. As a possible direction for future work, EVADE could be extended to partially observable domains without any problem decomposition.

\bibliographystyle{named}
\bibliography{references}

\begin{thebibliography}{}

\bibitem[\protect\citeauthoryear{Amato and Oliehoek}{2015}]{amato2015scalable}
Christopher Amato and Frans~A Oliehoek.
\newblock Scalable planning and learning for multiagent pomdps.
\newblock In {\em Proceedings of the Twenty-Ninth AAAI Conference on Artificial
  Intelligence}, pages 1995--2002. AAAI Press, 2015.

\bibitem[\protect\citeauthoryear{Anthony \bgroup \em et al.\egroup
  }{2017}]{thinkingfastandslow}
Thomas Anthony, Zheng Tian, and David Barber.
\newblock Thinking fast and slow with deep learning and tree search.
\newblock In {\em Advances in Neural Information Processing Systems}, pages
  5366--5376, 2017.

\bibitem[\protect\citeauthoryear{Bai \bgroup \em et al.\egroup
  }{2014}]{bai2014thompson}
Aijun Bai, Feng Wu, Zongzhang Zhang, and Xiaoping Chen.
\newblock Thompson sampling based monte-carlo planning in pomdps.
\newblock In {\em Proceedings of the Twenty-Fourth International Conferenc on
  International Conference on Automated Planning and Scheduling}, pages 29--37.
  AAAI Press, 2014.

\bibitem[\protect\citeauthoryear{Bellman}{1957}]{bellman1957}
Richard Bellman.
\newblock {\em Dynamic Programming}.
\newblock Princeton University Press, Princeton, NJ, USA, 1 edition, 1957.

\bibitem[\protect\citeauthoryear{Belzner and
  Gabor}{2017a}]{belzner2017scalable}
Lenz Belzner and Thomas Gabor.
\newblock Scalable multiagent coordination with distributed online open loop
  planning.
\newblock {\em arXiv preprint arXiv:1702.07544}, 2017.

\bibitem[\protect\citeauthoryear{Belzner and Gabor}{2017b}]{belzner2017stacked}
Lenz Belzner and Thomas Gabor.
\newblock Stacked thompson bandits.
\newblock In {\em Proceedings of the 3rd International Workshop on Software
  Engineering for Smart Cyber-Physical Systems}, pages 18--21. IEEE Press,
  2017.

\bibitem[\protect\citeauthoryear{Belzner \bgroup \em et al.\egroup
  }{2015}]{belzner2015onplan}
Lenz Belzner, Rolf Hennicker, and Martin Wirsing.
\newblock Onplan: A framework for simulation-based online planning.
\newblock In {\em International Workshop on Formal Aspects of Component
  Software}, pages 1--30. Springer, 2015.

\bibitem[\protect\citeauthoryear{Boutilier}{1996}]{boutilier1996planning}
Craig Boutilier.
\newblock Planning, learning and coordination in multiagent decision processes.
\newblock In {\em Proceedings of the 6th conference on Theoretical aspects of
  rationality and knowledge}, pages 195--210. Morgan Kaufmann Publishers Inc.,
  1996.

\bibitem[\protect\citeauthoryear{Bubeck and Munos}{2010}]{bubeck2010open}
S~Bubeck and R~Munos.
\newblock Open loop optimistic planning.
\newblock In {\em Conference on Learning Theory}, 2010.

\bibitem[\protect\citeauthoryear{Bu{\c{s}}oniu \bgroup \em et al.\egroup
  }{2010}]{bucsoniu2010multi}
Lucian Bu{\c{s}}oniu, Robert Babu{\v{s}}ka, and Bart De~Schutter.
\newblock Multi-agent reinforcement learning: An overview.
\newblock In {\em Innovations in multi-agent systems and applications-1}, pages
  183--221. Springer, 2010.

\bibitem[\protect\citeauthoryear{Cassandra and
  Kaelbling}{2016}]{cassandra2016learning}
Anthony~R Cassandra and Leslie~Pack Kaelbling.
\newblock Learning policies for partially observable environments: Scaling up.
\newblock In {\em Machine Learning Proceedings 1995: Proceedings of the Twelfth
  International Conference on Machine Learning, Tahoe City, California, July
  9-12 1995}, page 362. Morgan Kaufmann, 2016.

\bibitem[\protect\citeauthoryear{Chapelle and Li}{2011}]{chapelle2011empirical}
Olivier Chapelle and Lihong Li.
\newblock An empirical evaluation of thompson sampling.
\newblock In {\em Advances in neural information processing systems}, pages
  2249--2257, 2011.

\bibitem[\protect\citeauthoryear{Chaslot}{2010}]{chaslot2010monte}
Guillaume Chaslot.
\newblock Monte-carlo tree search.
\newblock {\em Maastricht: Universiteit Maastricht}, 2010.

\bibitem[\protect\citeauthoryear{Claes \bgroup \em et al.\egroup
  }{2017}]{claes2017decentralised}
Daniel Claes, Frans Oliehoek, Hendrik Baier, and Karl Tuyls.
\newblock Decentralised online planning for multi-robot warehouse
  commissioning.
\newblock In {\em Proceedings of the 16th Conference on Autonomous Agents and
  MultiAgent Systems}, pages 492--500. International Foundation for Autonomous
  Agents and Multiagent Systems, 2017.

\bibitem[\protect\citeauthoryear{Clevert \bgroup \em et al.\egroup
  }{2015}]{clevert2015fast}
Djork-Arn{\'e} Clevert, Thomas Unterthiner, and Sepp Hochreiter.
\newblock Fast and accurate deep network learning by exponential linear units
  (elus).
\newblock {\em CoRR}, abs/1511.07289, 2015.

\bibitem[\protect\citeauthoryear{Emery-Montemerlo \bgroup \em et al.\egroup
  }{2004}]{emery2004approximate}
Rosemary Emery-Montemerlo, Geoff Gordon, Jeff Schneider, and Sebastian Thrun.
\newblock Approximate solutions for partially observable stochastic games with
  common payoffs.
\newblock In {\em Proceedings of the Third International Joint Conference on
  Autonomous Agents and Multiagent Systems-Volume 1}, pages 136--143. IEEE
  Computer Society, 2004.

\bibitem[\protect\citeauthoryear{Evans}{1984}]{evans1984heuristic}
Jonathan St~BT Evans.
\newblock Heuristic and analytic processes in reasoning.
\newblock {\em British Journal of Psychology}, 75(4):451--468, 1984.

\bibitem[\protect\citeauthoryear{Foerster \bgroup \em et al.\egroup
  }{2016}]{foerster2016learning}
Jakob Foerster, Yannis~M Assael, Nando de~Freitas, and Shimon Whiteson.
\newblock Learning to communicate with deep multi-agent reinforcement learning.
\newblock In {\em Advances in Neural Information Processing Systems}, pages
  2137--2145, 2016.

\bibitem[\protect\citeauthoryear{Hester and Stone}{2013}]{hester2013texplore}
Todd Hester and Peter Stone.
\newblock Texplore: real-time sample-efficient reinforcement learning for
  robots.
\newblock {\em Machine learning}, 90(3):385--429, 2013.

\bibitem[\protect\citeauthoryear{Honda and
  Takemura}{2014}]{honda2014optimality}
Junya Honda and Akimichi Takemura.
\newblock Optimality of thompson sampling for gaussian bandits depends on
  priors.
\newblock In {\em Artificial Intelligence and Statistics}, pages 375--383,
  2014.

\bibitem[\protect\citeauthoryear{Howard}{1961}]{beranek1961ronald}
Ronald~A. Howard.
\newblock {\em Dynamic Programming and Markov Processes}.
\newblock The MIT Press, 1961.

\bibitem[\protect\citeauthoryear{Kahneman}{2003}]{kahneman2003maps}
Daniel Kahneman.
\newblock Maps of bounded rationality: Psychology for behavioral economics.
\newblock {\em The American economic review}, 93(5):1449--1475, 2003.

\bibitem[\protect\citeauthoryear{Kocsis and
  Szepesv{\'a}ri}{2006}]{kocsis2006bandit}
Levente Kocsis and Csaba Szepesv{\'a}ri.
\newblock Bandit based monte-carlo planning.
\newblock In {\em ECML}, volume~6, pages 282--293. Springer, 2006.

\bibitem[\protect\citeauthoryear{Mnih \bgroup \em et al.\egroup
  }{2013}]{mnih2013playing}
Volodymyr Mnih, Koray Kavukcuoglu, David Silver, Alex Graves, Ioannis
  Antonoglou, Daan Wierstra, and Martin Riedmiller.
\newblock Playing atari with deep reinforcement learning.
\newblock In {\em NIPS Deep Learning Workshop}. 2013.

\bibitem[\protect\citeauthoryear{Mnih \bgroup \em et al.\egroup
  }{2015}]{mnih2015human}
Volodymyr Mnih, Koray Kavukcuoglu, David Silver, Andrei~A Rusu, Joel Veness,
  Marc~G Bellemare, Alex Graves, Martin Riedmiller, Andreas~K Fidjeland, Georg
  Ostrovski, et~al.
\newblock Human-level control through deep reinforcement learning.
\newblock {\em Nature}, 518(7540):529--533, 2015.

\bibitem[\protect\citeauthoryear{Mnih \bgroup \em et al.\egroup
  }{2016}]{mnih2016asynchronous}
Volodymyr Mnih, Adria~Puigdomenech Badia, Mehdi Mirza, Alex Graves, Timothy
  Lillicrap, Tim Harley, David Silver, and Koray Kavukcuoglu.
\newblock Asynchronous methods for deep reinforcement learning.
\newblock In {\em International Conference on Machine Learning}, pages
  1928--1937, 2016.

\bibitem[\protect\citeauthoryear{Nair \bgroup \em et al.\egroup
  }{2015}]{nair2015massively}
Arun Nair, Praveen Srinivasan, Sam Blackwell, Cagdas Alcicek, Rory Fearon,
  Alessandro De~Maria, Vedavyas Panneershelvam, Mustafa Suleyman, Charles
  Beattie, Stig Petersen, et~al.
\newblock Massively parallel methods for deep reinforcement learning.
\newblock {\em arXiv preprint arXiv:1507.04296}, 2015.

\bibitem[\protect\citeauthoryear{Oliehoek and
  Amato}{2016}]{oliehoek2016concise}
Frans~A Oliehoek and Christopher Amato.
\newblock {\em A concise introduction to decentralized POMDPs}.
\newblock Springer, 2016.

\bibitem[\protect\citeauthoryear{Oliehoek \bgroup \em et al.\egroup
  }{2008a}]{oliehoek2008cross}
Frans~A Oliehoek, Julian~FP Kooij, and Nikos Vlassis.
\newblock The cross-entropy method for policy search in decentralized pomdps.
\newblock {\em Informatica}, 32(4), 2008.

\bibitem[\protect\citeauthoryear{Oliehoek \bgroup \em et al.\egroup
  }{2008b}]{oliehoek2008optimal}
Frans~A Oliehoek, Matthijs~TJ Spaan, and Nikos Vlassis.
\newblock Optimal and approximate q-value functions for decentralized pomdps.
\newblock {\em Journal of Artificial Intelligence Research}, 32:289--353, 2008.

\bibitem[\protect\citeauthoryear{Paquet \bgroup \em et al.\egroup
  }{2006}]{paquet2006hybrid}
S{\'e}bastien Paquet, Brahim Chaib-draa, and St{\'e}phane Ross.
\newblock Hybrid pomdp algorithms.
\newblock In {\em Proceedings of The Workshop on Multi-Agent Sequential
  Decision Making in Uncertain Domains (MSDM-06)}, pages 133--147, 2006.

\bibitem[\protect\citeauthoryear{Perez~Liebana \bgroup \em et al.\egroup
  }{2015}]{perez2015open}
Diego Perez~Liebana, Jens Dieskau, Martin Hunermund, Sanaz Mostaghim, and Simon
  Lucas.
\newblock Open loop search for general video game playing.
\newblock In {\em Proceedings of the 2015 Annual Conference on Genetic and
  Evolutionary Computation}, pages 337--344. ACM, 2015.

\bibitem[\protect\citeauthoryear{Puterman}{2014}]{puterman2014markov}
Martin~L Puterman.
\newblock {\em Markov decision processes: discrete stochastic dynamic
  programming}.
\newblock John Wiley \& Sons, 2014.

\bibitem[\protect\citeauthoryear{Ross \bgroup \em et al.\egroup
  }{2007}]{ross2007aems}
St{\'e}phane Ross, Brahim Chaib-Draa, et~al.
\newblock Aems: An anytime online search algorithm for approximate policy
  refinement in large pomdps.
\newblock In {\em IJCAI}, pages 2592--2598, 2007.

\bibitem[\protect\citeauthoryear{Silver and Veness}{2010}]{silver2010monte}
David Silver and Joel Veness.
\newblock Monte-carlo planning in large pomdps.
\newblock In {\em Advances in neural information processing systems}, pages
  2164--2172, 2010.

\bibitem[\protect\citeauthoryear{Silver \bgroup \em et al.\egroup
  }{2016}]{silver2016mastering}
David Silver, Aja Huang, Chris~J Maddison, Arthur Guez, Laurent Sifre, George
  Van Den~Driessche, Julian Schrittwieser, Ioannis Antonoglou, Veda
  Panneershelvam, Marc Lanctot, et~al.
\newblock Mastering the game of go with deep neural networks and tree search.
\newblock {\em Nature}, 529(7587):484--489, 2016.

\bibitem[\protect\citeauthoryear{Silver \bgroup \em et al.\egroup
  }{2017}]{silver2017mastering}
David Silver, Julian Schrittwieser, Karen Simonyan, Ioannis Antonoglou, Aja
  Huang, Arthur Guez, Thomas Hubert, Lucas Baker, Matthew Lai, Adrian Bolton,
  et~al.
\newblock Mastering the game of go without human knowledge.
\newblock {\em Nature}, 550(7676):354--359, 2017.

\bibitem[\protect\citeauthoryear{Spaan \bgroup \em et al.\egroup
  }{2011}]{spaan2011scaling}
Matthijs~TJ Spaan, Frans~A Oliehoek, and Christopher Amato.
\newblock Scaling up optimal heuristic search in dec-pomdps via incremental
  expansion.
\newblock In {\em Proceedings of the Twenty-Second international joint
  conference on Artificial Intelligence-Volume Volume Three}, pages 2027--2032.
  AAAI Press, 2011.

\bibitem[\protect\citeauthoryear{Sutton and
  Barto}{1998}]{sutton1998introduction}
Richard~S Sutton and Andrew~G Barto.
\newblock {\em Introduction to reinforcement learning}, volume 135.
\newblock MIT Press Cambridge, 1998.

\bibitem[\protect\citeauthoryear{Sutton}{1988}]{sutton1988learning}
Richard~S Sutton.
\newblock Learning to predict by the methods of temporal differences.
\newblock {\em Machine learning}, 3(1):9--44, 1988.

\bibitem[\protect\citeauthoryear{Szer \bgroup \em et al.\egroup
  }{2005}]{szer2005maa}
Daniel Szer, Francois Charpillet, and Shlomo Zilberstein.
\newblock Maa*: A heuristic search algorithm for solving decentralized pomdps.
\newblock In {\em 21st Conference on Uncertainty in Artificial
  Intelligence-UAI'2005}, 2005.

\bibitem[\protect\citeauthoryear{Tampuu \bgroup \em et al.\egroup
  }{2017}]{tampuu2017multiagent}
Ardi Tampuu, Tambet Matiisen, Dorian Kodelja, Ilya Kuzovkin, Kristjan Korjus,
  Juhan Aru, Jaan Aru, and Raul Vicente.
\newblock Multiagent cooperation and competition with deep reinforcement
  learning.
\newblock {\em PloS one}, 12(4):e0172395, 2017.

\bibitem[\protect\citeauthoryear{Tan}{1997}]{tan1997multi}
Ming Tan.
\newblock Multi-agent reinforcement learning: independent vs. cooperative
  agents.
\newblock In {\em Readings in agents}, pages 487--494. Morgan Kaufmann
  Publishers Inc., 1997.

\bibitem[\protect\citeauthoryear{Thompson}{1933}]{thompson1933likelihood}
William~R Thompson.
\newblock On the likelihood that one unknown probability exceeds another in
  view of the evidence of two samples.
\newblock {\em Biometrika}, 25(3/4):285--294, 1933.

\bibitem[\protect\citeauthoryear{Weinstein and
  Littman}{2013}]{weinstein2013open}
Ari Weinstein and Michael~L Littman.
\newblock Open-loop planning in large-scale stochastic domains.
\newblock In {\em AAAI}, 2013.

\bibitem[\protect\citeauthoryear{Wu \bgroup \em et al.\egroup
  }{2009}]{wu2009multi}
Feng Wu, Shlomo Zilberstein, and Xiaoping Chen.
\newblock Multi-agent online planning with communication.
\newblock In {\em Nineteenth International Conference on Automated Planning and
  Scheduling}, 2009.

\end{thebibliography}

\end{document}